\documentclass[a4paper,fleqn,usenatbib]{mnras}

\usepackage{mathptmx}

\usepackage[T1]{fontenc}
\usepackage{ae,aecompl}

\usepackage{graphicx}	
\usepackage{amsmath}	
\usepackage{amssymb}	

\usepackage{color}

\newcommand{\obj}{\hbox{KIC\,2831097}}
\newcommand{\kep}{{\it Kepler}}
\newcommand{\oc}{\hbox{\textit{O--C}}}
\newcommand{\cd}{\,d$^{-1}$}

\title[An RRc binary candidate]{KIC\,2831097 -- A 2-year orbital-period RR Lyrae binary candidate}

\author[\'A. S\'odor et al.]{
\'A. S\'odor,$^{1}$\thanks{\tt{sodor@konkoly.hu}}
M. Skarka,$^{1}$\thanks{\tt{marek.skarka@csfk.mta.hu}}
J. Li\v{s}ka,$^{2}$\thanks{\tt{jiriliska@post.cz}}
and Zs. Bogn\'ar$^{1}$\thanks{\tt{bognar@konkoly.hu}}
\\
$^{1}$Konkoly Observatory, MTA Research Centre for Astronomy and Earth Sciences, Konkoly Thege \'ut 15-17, H--1121 Budapest, Hungary\\
$^{2}$Department of Theoretical Physics and Astrophysics, Masaryk University, Kotl\'a\u rsk\'a 2, CZ-611 37 Brno, Czech Republic
}

\date{Accepted XXX. Received YYY; in original form ZZZ}

\pubyear{2016}

\begin{document}
\label{firstpage}
\pagerange{\pageref{firstpage}--\pageref{lastpage}}
\maketitle

\begin{abstract}
We report the discovery of a new \kep\/ first-overtone RR Lyrae pulsator, \obj. The pulsation shows large, 0.1\,d amplitude, systematic phase variations that can be interpreted as light travel-time effect caused by orbital motion in a binary system, superimposed on a linear pulsation-period decrease. The assumed eccentric ($e=0.47$) orbit with the period of $\approx 2$\,yr is the shortest among the non-eclipsing RR Lyrae binary candidates. The binary model gives a lowest estimate for the mass of the companion of 8.4\,$\mathfrak{M}_{\odot}$, that places it among black hole candidates. Beside the first-overtone pulsation, numerous additional non-radial pulsation frequencies were also identified. We detected an $\approx 47$-d Blazhko-like irregular light-curve modulation.
\end{abstract}

\begin{keywords}
	stars: variables: RR Lyrae -- 
	binaries: general -- 
	methods: data analysis -- 
	techniques: photometric --
	stars: individual: KIC 2831097
\end{keywords}



\section{Introduction}

Past decades showed that the pulsation period of many RR Lyrae stars undergo long-term variations that cannot be explained by simple evolutionary effects. Beside abrupt changes \citep[see e.g.][]{lacluyze2004,sodor2007}, many stars show well documented cyclic years-to-decades-long period variations in the Galactic field \citep[e.g.][]{firmanyuk1982}, but mainly in globular clusters \citep[see e.g.][]{jurcsik2011,szeidl2011}. These variations often resemble the Light Travel-time Effect (LiTE), the consequence of orbital motion in combination with the finite speed of light.

The investigation of period changes is the most efficient method for revealing RR Lyrae binary candidates, because typical RR Lyrae characteristics, mainly their evolutionary stadium and luminosity, disadvantage other available methods \citep[see e.g. discussions by][]{richmond2011,skarka2016}. Recently, several studies based on time-delay analysis introducing a few tens of binary candidates were published \citep[][]{li2014,guggenberger2015,hajdu2015,liska2016b,dePonthiere2016JAVSO..44...18D}. The most promising star, showing distinct, well-defined cyclic changes, is TU~UMa \citep[for a detailed study see][]{liska2016a}. Unfortunately, none of the candidates has yet been fully confirmed spectroscopically\footnote{An ongoing spectroscopic campaign on several candidates takes place right now \citep{guggenberger2016}.}, and the only confirmed RR Lyrae-like variable in eclipsing system turned out not to be a classical RR Lyrae, but the product of evolution in a close binary system \citep{pietrzynski2012,smolec2013}. Therefore, firm identification of an RR Lyrae in binary system supported by spectroscopic observations is still of extremely high importance favouring candidates with the shortest proposed, year-long, orbital periods. In this research note, we present one such candidate.

We identified \obj\ as a previously unknown first-overtone RR Lyrae-type variable (RRc)\footnote{After identification, we found a Planet Hunters blog entry from Oct. 2012 on the suspected RRc nature at {\tt http://bit.ly/28Ip1Nc}}. As detailed light-curve analysis of only four \kep\ RRc stars has been published to date \citep{moskrrc}, this itself makes \obj\ an interesting target. However, our analysis revealed that this star might be a binary with the shortest know orbital period among non-eclipsing RR Lyrae binary candidates.

In this research note, we analyse the \oc\/ variations in the context of hypothetical binarity, and present a concise analysis of the pulsation of \obj. We also present pro and con arguments of the binary explanation, and discuss the further observations necessary to confirm or reject the binary hypothesis.

\section{The data}

KIC\,2831097 ($\alpha_\mathrm{2000} = +19^\mathrm{h}02^\mathrm{m}06^\mathrm{s}$, $\delta_\mathrm{2000} = +38^\circ04\arcmin41\arcsec$, $K_p = 14.5$\,mag) was observed during the original programme of the \kep\/ space telescope \citep{Borucki2010Sci...327..977B,Koch2010ApJ...713L..79K} between JD\,2\,455\,003 and JD\,2\,456\,424 (Q2\,--\,Q17). The duty cycle of the 1421-d long-cadence (29.4-min time resolution; \citealt{Jenkins2010ApJ...713L.120J}) time string is $\approx90$ per~cent. For this analysis, we used the Pre-search Data Conditioning Single APerture (PDC SAP; \citealt{Smith2012PASP..124.1000S}) flux data downloaded from the MAST\footnote{\tt{https://archive.stsci.edu/kepler/}} database in Feb. 2016.

Long-term instrumental fluctuations were removed by normalising the contiguous data blocks with splines fitted to the averages of 5-d segments. Only obvious outliers were removed at this stage. We performed a second pass of outlier elimination and trend filtering based on sigma-clipping and splines fitted to the residuals after subtracting a preliminary Fourier solution from the data. The light curve pre-processed this way contains 62\,267 data points.

\begin{figure}
  \includegraphics[width=\columnwidth]{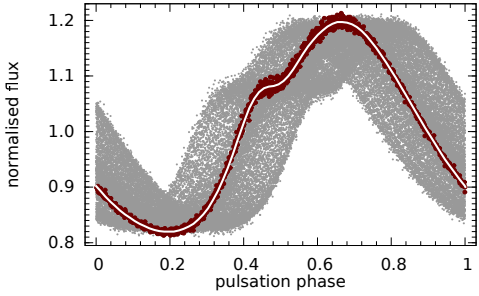}
  \caption{The light curve of \obj\ folded with the pulsation period. Dark red dots highlight data from a short time interval of 30\,d that was fitted with harmonics of the pulsation frequency (template curve -- white continuous curve).\label{fig:lc}}
\end{figure}

\section{Pulsation and phase variations}

The light curve phased with the average of the main pulsation period determined for the whole data set (0.37705\,d) is presented in the left-hand panel of Fig.~\ref{fig:lc}. The diagram shows strong phase variations, while the amplitude remained nearly constant. We begin our analysis with disentangling the large-scale phase variations and the pulsation.

To describe the shape of the pulsation light curve unaffected by phase variations, we fitted 9 harmonics of the main pulsation frequency to a 30-d section of the data starting at JD\,2\,455\,243, where phase variation is negligible (template curve; see Fig.~\ref{fig:lc}).

We determined the phase variations by the \oc\/ method. For the calculated times of maxima ($C$) we used the ephemeris
$$ T_\mathrm{max} = \mathrm{BJD}\,2\,455\,002.5528 + 0.3770501\,\mathrm{d}\cdot E.$$
The observed times of maxima ($O$) were determined by fitting the phase shift and an amplitude scaling factor of the pulsation template curve to overlapping 3-d data segments. The \oc\/ curve, shown in the top panel of Fig.~\ref{fig:orb}, contains 1310 data points with 1-d sampling. The bottom panel shows the relative variations in the pulsation amplitude (the mean peak-to-peak pulsation amplitude is 0.33 mag).

\section{Orbital solution}

\begin{figure}
	\centering
	\includegraphics[width=0.9\columnwidth]{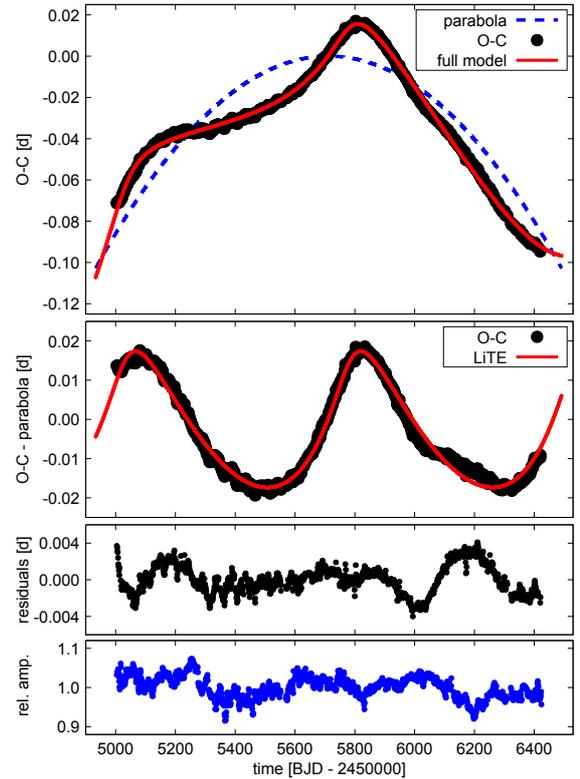}
   \caption{The \oc\/ diagram of \obj. Top: data fitted with constant period decrease (blue dashed curve), and the superimposed orbital light-delay variations (red continuous curve). Second panel: the data after the parabolic variation is subtracted, with the fitted orbital variation (continuous curve). Third panel: final residuals. Bottom: relative pulsation amplitude.\label{fig:orb}}
\end{figure}

The long-term variations in the \oc\/ diagram, shown in the top panel of Fig.~\ref{fig:orb}, can be interpreted as LiTE, as the pulsating star orbits a companion. The periodic orbital \oc\/ variations are superimposed on a constant period decrease -- corresponding to a parabola. For modelling the observed changes, we used a code introduced in \citet{liska2016a}, where the parabola and the orbital motion in the binary system are simultaneously solved. The best model (Fig.~\ref{fig:orb}) was found using a non-linear least-squares method. Afterwards, uncertainties of individual parameters were determined statistically by bootstrap-resampling (5\,000 repetition of the calculation with re-sampled datasets). The final pulsation and orbital parameters are summarised in Table~\ref{Tab:LiTEparam}.

\begin{table}
  \begin{minipage}{\columnwidth}
    \begin{center}
    \caption{Pulsation and orbital parameters for binary hypothesis. Errors are given in parentheses in the unit of the last digits.\label{Tab:LiTEparam}}
      \begin{tabular}{lcr@{.}l}
        \hline
        Parameter$^{*}$               & unit                   & \multicolumn{2}{c}{quantity} \\
        \hline
        $\dot{P}_{\rm puls}$          & $10^{-7}$\,d\,d$^{-1}$ & $-1$&277(2) \\
        $\beta$                       & d\,Myr$^{-1}$          & $-46$&66(8)\\
        $P_{\rm orbit}$               & d                      & 752&9(1.1)\\
        $T_{0}$                       & BJD                    & 2\,455\,780&(2)\\
        $e$                           &                        & 0&474(6)\\
        $\omega$                      & $^{\circ}$             & 52&0(8)\\
        $a_{1}\sin i$                 & au                     & 3&143(9)\\
        $A$                           & light day              & 0&01815(5)\\
        $f(\mathfrak{M})$             & $\mathfrak{M}_{\odot}$ & 7&31(7)\\
        $\mathfrak{M}_{\rm 2, min}$   & $\mathfrak{M}_{\odot}$ & 8&39(7)\\
        $K_{1}$                       &  km\,s$^{-1}$          & 51&6(3)\\
        $\chi_{\rm R}^2$              &                        & 1&01(4)\\
        \hline
      \end{tabular}
    \end{center}
  {\scriptsize $^{*}$ 
  $\dot{P}_{\rm puls} = \beta$ -- rate of linear pulsation-period changes, 
  $P_{\rm orbit}$ -- orbital period, 
  $T_{0}$ -- time of periastron passage, 
  $e$ -- numerical eccentricity, 
  $\omega$ -- argument of periastron, 
  $a_{1}\sin i$ -- projection of semi-major axis of primary component $a_{1}$ according to the inclination of the orbit $i$,
  $A$ -- $a_{1}\sin i$ in light days (semi-amplitude of LiTE $A_{\rm LiTE}$ can be calculated as $A_{\rm LiTE} = A\,\sqrt{1-e^{2}\,\cos^{2}\omega}$ ), 
  $f(\mathfrak{M})$ -- mass function, 
  $\mathfrak{M}_{\rm 2, min}$ -- the lowest mass of the secondary component, the value was calculated for inclination angle $i=90^{\circ}$ and adopted mass of primary $\mathfrak{M}_{1} = 0.6$\,$\mathfrak{M}_{\odot}$, 
  $K_{1}$ -- semi-amplitude of RV changes of the pulsating star, 
  $\chi_{\rm R}^2$ -- normalised value of $\chi^{2}$, where $\chi^{2}_{\rm R} = \chi^{2}/(N-g)$ for number of used values $N$ and number of free (fitted) parameters $g$ (LiTE + parabola $g=8$).} 
  \end{minipage}
\end{table}

\section{Fourier analysis of the pulsation}

According to the binary interpretation, the periodic component of the \oc\/ variations (Fig.~\ref{fig:orb}, second panel) is not intrinsic to the pulsator. Therefore, the orbital phase variations have to be eliminated from the data before performing the pulsational analysis. It was achieved by applying a time-transformation according to the orbital solution; we subtracted the orbital component of the \oc\/ solution from the times of the observations.

The main pulsation still showed the large-scale parabolic phase-variation trend after the time-transformation. Thus, we applied the time-dependent prewhitening method described by \citet[][appendix A]{moskrrc}, taking into account only the parabolic phase variation of the pulsation, but neglecting the short-term phase fluctuations and amplitude variations.

\begin{figure*}
  \includegraphics[width=180mm]{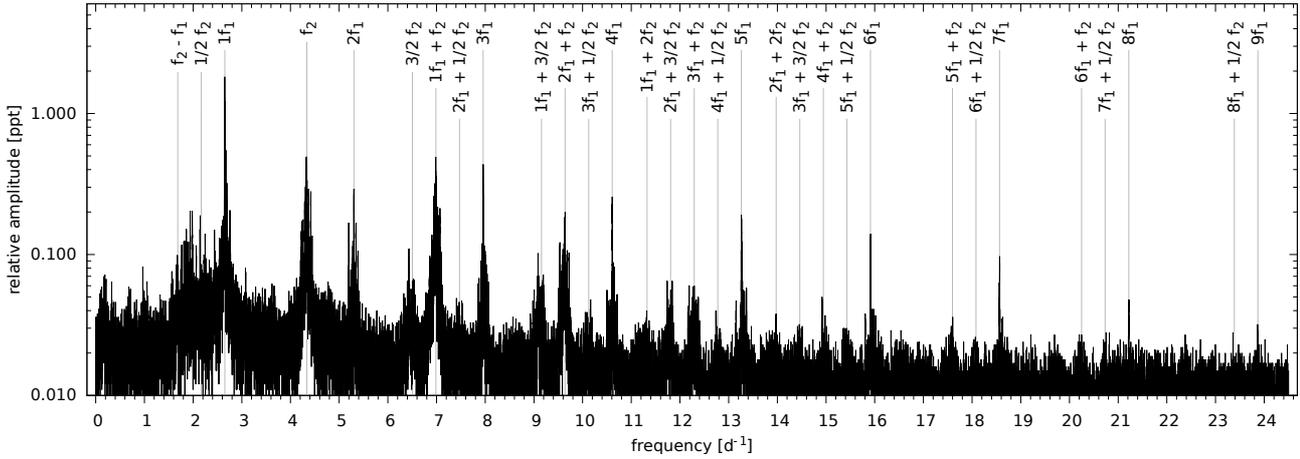}
  \caption{Fourier amplitude spectrum of the time-transformed light curve of \obj\ prewhitened with the pulsation frequency and its harmonics.\label{fig:sp1}}
\end{figure*}

\begin{figure}
  \includegraphics[width=\columnwidth]{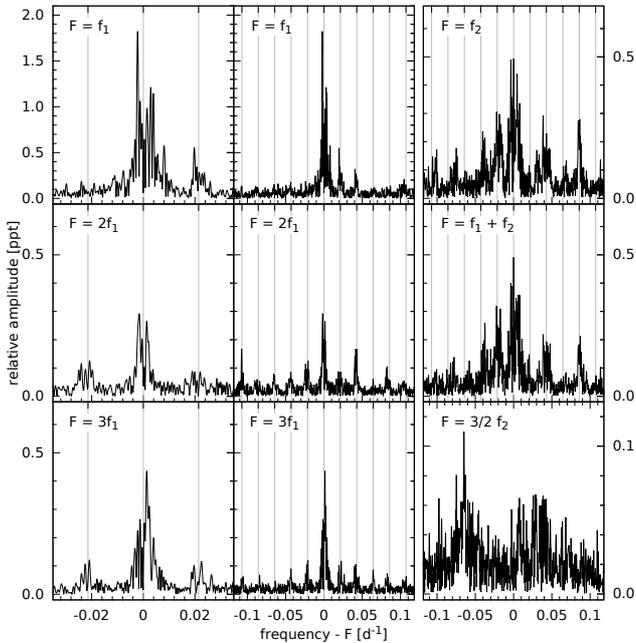}
  \caption{The vicinity of some of the most prominent peaks in the Fourier spectrum of the time-transformed light curve of \obj\ after prewhitened with the main pulsation frequency and its harmonics. Left-hand and middle columns show the first three pulsation harmonics ($kf_1$) over different frequency ranges. The right-hand column shows the vicinity of $f_2 = 1.63 f_1$, one of its linear combinations with $f_1$, and the $3/2 f_2$ subharmonic. Vertical grey lines indicate $\pm jf_\mathrm{m}$ separations, where $f_\mathrm{m}=0.0214$\cd.
  \label{fig:sp2}}
\end{figure}

The Fourier spectrum of the prewhitened data treated this way is presented in Fig.~\ref{fig:sp1}. The vicinity of the most prominent peaks are shown expanded in Fig.~\ref{fig:sp2}. The strongest residual peaks are around the pulsation harmonics, within less than 0.01\cd (Fig.~\ref{fig:sp2}, left-hand column). These are not sharp, equidistant peaks, their structure and spacing differs from harmonic to harmonic. These side peaks are related to long-term ($> 100$\,d), irregular amplitude variations, as well as residual phase variations after the removal of the orbital solution and the parabolic trend. The different residual structure around the different pulsation harmonics also suggest irregular long-term variations in the light curve shape.

Looking at the broader vicinity of the pulsation harmonics (Fig.~\ref{fig:sp2}, center column), many broadened features are visible, which are regularly spaced around the central peak by $\pm jf_\mathrm{m}$ separations up to $j=5$, where $f_\mathrm{m}=0.0214$\cd. These peaks resemble the expression of the Blazhko-modulation in the Fourier spectrum.

The Fourier spectrum shows a secondary pulsation frequency at $f_2=4.332$\cd\ (Fig.~\ref{fig:sp2} top right panel). The frequency ratio to the main pulsation frequency is $f_1/f_2=0.612$. Additional frequencies with very similar frequency ratios ($f_1/f_2=0.61$\,--\,0.63) were already found in RRc stars \citep[][and references therein]{jurcsik15m3,Netzel2015MNRAS.447.1173N,moskrrc}, particularly in all four previously studied \kep\ RRc stars \citep{moskrrc}. Similarly to other RRc stars, the peak in the Fourier spectrum at $f_2$ is much broader than the harmonics of the main pulsation, indicating that this pulsation mode is less stable than the dominant one. As well as the pulsation harmonics, $f_2$ is also surrounded by broadened side peaks, however, these are generally broader and their separation from the central peak is not as regular as those around the harmonics of the main pulsation.

The three most important independent frequencies are listed in Table~\ref{tbl:freq}. Note that the listed frequencies are approximate mean values only, since each of them changed significantly during the observations. The detailed investigation of these frequency variations is beyond the scope of the present research note.

Linear combinations of $kf_1+f_2$ also appear in the Fourier spectrum, as well as half-integer multiples of $f_2$, and their linear combinations with $f_1$ (Figs.~\ref{fig:sp1} and \ref{fig:sp2}). Similarly to $f_2$, these combination signals do not appear as sharp peaks in the Fourier spectrum, but they are broadened, sometimes even more than $f_2$ itself.

\begin{figure}
  \includegraphics[width=\columnwidth]{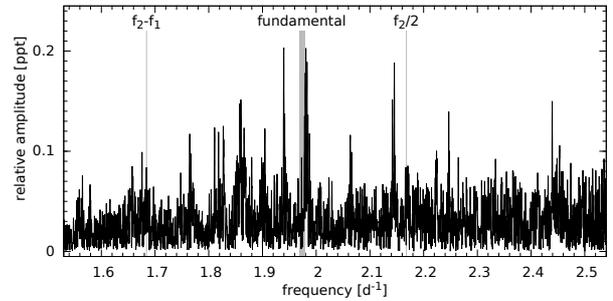}
  \caption{Additional peaks in the Fourier spectrum of the time-transformed light curve of \obj\ prewhitened with the main pulsation frequency and its harmonics.\label{fig:sp3}}
\end{figure}

\begin{table}
  \begin{center}
    \caption{The three most important frequencies found in the time-transformed data.\label{tbl:freq}}
    \begin{tabular}{lll}
      \hline
      Component      & frequency [\cd] & period [d] \\
      \hline
      $f_1$          & 2.652        & \ 0.377 \\
      $f_2$          & 4.332          & \ 0.231 \\
      $f_\mathrm{m}$ & 0.021          & 47.0\\
      \hline
    \end{tabular}
  \end{center}
\end{table}

We can find additional peaks independent of $f_1$, $f_2$ and $f_\mathrm{m}$ in the 1.5\,--\,2.5\cd\ range (Fig.~\ref{fig:sp3}). This is also the vicinity where the fundamental radial pulsation frequency ($f_0$) is expected. In Fig.~\ref{fig:sp3}, a grey band indicates the range of $f_0/f_1$\,=\,0.742\,--\,0.746, where $f_0$ is expected according to \citet[][fig.~4]{Soszynski2011AcA....61....1S}. There is a broad feature at $f_\mathrm{x1}=1.983$\cd, corresponding to $f_\mathrm{x1}/f_1=0.748$, and another marginally significant peak at $f_\mathrm{x2}=1.973$\cd ($f_\mathrm{x2}/f_1=0.744$). Any of these two might be the fundamental radial mode, but it is also possible that neither of them is. However, at most one peak may correspond to the fundamental mode, thus, the rest of the independent frequencies in the 1.5\,--\,2.5\cd\ range must correspond to non-radial pulsation modes. \cite{moskrrc} also found non-radial modes in \kep\ RRc stars.

\section{Discussion and Summary}

\subsection{Pulsation}
\label{sect:pulsdisc}
KIC\,2831097 appears to be a typical first-overtone RR Lyrae pulsator. It shows a wide variety of pulsational phenomena that were already found in the other four investigated \kep\ RRc stars by \cite{moskrrc}. These are: 
\begin{itemize}
 \item A secondary pulsation frequency, $f_2$ with $f_1/f_2=0.612$ ratio to the first-overtone pulsation frequency, $f_1$.
 \item Half-integer multiples (subharmonics) of $f_2$.
 \item Linear combinations of $f_1$ with $f_2$ and its subharmonics.
 \item All peaks are broadened in the periodogram of the time-transformed data.
 \item Strong phase variations and weak, irregular amplitude changes on the order of $\pm$10 per cent.
 \item Additional, non-radial pulsation modes.
 \item Many Blazhko-modulation-like frequency components around the harmonics of $f_1$.
\end{itemize}

Blazhko-modulation-like side peaks around the pulsation harmonics ($kf_1+jf_\mathrm{m}$) can be found up to $j=5$. These peaks are equidistant, but broadened. The broadening indicates that the modulation is not strictly periodic. The wiggles visible in the residual \oc\/ variations (3rd panel of Fig.~\ref{fig:orb}), especially in the first half of the data, show an $\approx47$-d periodicity corresponding to $f_\mathrm{m}$.

An interesting feature of the Fourier spectrum of \obj\ is the periodic pattern visible in Fig.~\ref{fig:sp1}. The overall distribution of the peaks is quite regular: 1\,--\,4 broadened peaks precede each pulsation harmonic with equidistant spacing of $f_1-1/2 f_2 = 0.48$\,\cd. This regular spacing has a simple arithmetic origin, as the linear-combination labels in the top part of the figure indicate.

\subsection{Binarity}

According to our orbital solution to the \oc\/ variations, the binary system has an eccentric orbit with quite high eccentricity of 0.47 and orbital period of 753\,d (2.06\,yr). This is the shortest known orbital period non-eclipsing RR Lyrae binary candidate \citep[see the up-to-date RRLyrBinCan database\footnote{http://rrlyrbincan.physics.muni.cz/},][]{liska2016b}. The orbital parameters and an adopted mass of the RR Lyrae component of $\mathfrak{M}_{1} = 0.6$\,$\mathfrak{M}_{\odot}$ predicts a very high mass for the possible companion; $\mathfrak{M}_{2} = 8.4$\,$\mathfrak{M}_{\odot}$, that places it among black hole candidates, due to evolutionary reasons.

Beside the high mass of the companion, the rate of the linear period change is also extraordinary. Parameter $\beta$ with $-46.7$\,d\,Myr$^{-1}$ is almost 62 times higher than for the star SW~Psc with record value of period shortening ($\beta = -0.756(61)$\,d\,Myr$^{-1}$) among the field RR Lyrae stars \citep{leborgne2007}.

All in all, there are several arguments both in favour and against the binary hypothesis.

\begin{itemize}
 \item [] Pro arguments are:
 \item [+] Almost two complete, repetitive orbital cycles can be followed in the \oc\/ diagram.
 \item [+] The phase variations are not accompanied by correlated amplitude variations that would suggest Blazhko modulation.
 \item [+] Deviations from the orbital \oc\/ curve appear to follow a different periodicity of around 900\,d (see Fig.~\ref{fig:orb})
 \vskip2mm
 \item []Con arguments are:
 \item [--] If \obj\ is a binary, the significant residual \oc\/ variations still require a different explanation.
 \item [--] Long-period RRc stars often show strong \oc\/ variations \citep{moskrrc}, even in a periodic fashion \citep{Derekas2004MNRAS.354..821D}.
 \item [--] This is the second long-pulsation-period RRc  binary candidate appear to orbit a black hole companion after BE~Dor \citep{Derekas2004MNRAS.354..821D}, which is at least suspicious.
\end{itemize}

The \kep\ data in itself is insufficient to decide the question of binarity. An important goal of this research note is to facilitate further observations on this object both spectroscopically and photometrically. We already initiated a ground-based photometric campaign to follow-up \oc\/ variations to extend the time base. Considering the relatively short predicted orbital period and large predicted gamma velocity amplitude, several short spectroscopic observing campaigns with 4-m class telescopes in the following 2\,--\,4 observing seasons must be decisive on this matter. Our new \oc\/ data will help timing these spectroscopic follow-up observations.

On the other hand, if \obj\ would prove to be a non-binary RR Lyrae star, it will be a warning that even single RR Lyrae can produce \oc\ curves that can be satisfactorily modelled with the combination of orbital light delay and linear period change, although the orbital parameters might be physically less plausible \citep[see also][]{Derekas2004MNRAS.354..821D}. This might be especially true for objects with less precise and less densely sampled \oc\ curves than the excellent \kep\/ data on our target.

\section*{Acknowledgements}

We are grateful to the anonymous referee for their insightful comments and suggestions that helped improving this research note.

The financial support of the Hungarian NKFIH Grants K-115709 and K-113117 are acknowledged. \'AS was supported by the J\'anos Bolyai Research Scholarship of the Hungarian Academy of Sciences. MS acknowledges the support of the postdoctoral fellowship programme of the Hungarian Academy of Sciences at the Konkoly Observatory as a host institution. This paper includes data collected by the Kepler mission. Funding for the Kepler mission is provided by the NASA Science Mission directorate. All of the data presented in this paper were obtained from the Mikulski Archive for Space Telescopes (MAST). 




\bibliographystyle{mnras}
\bibliography{sodor} 



%
%


\bsp	
\label{lastpage}
\end{document}